\documentclass[12pt,preprint]{aastex}
\usepackage{bbm}
\usepackage{mathrsfs}
\usepackage{amssymb,amsmath,float}
\usepackage{rotating}
\usepackage{color}

\newcommand\ha{\hat a}
\newcommand\had{\hat a^\dag}
\newcommand\hb{\hat b}
\newcommand\hbd{\hat b^\dag}
\newcommand\hc{\hat c}
\newcommand\hcd{\hat c^\dag}
\newcommand\hd{\hat d}
\newcommand\hdd{\hat d^\dag}
\newcommand\hD{\hat D}

\newcommand\hJ{\hat J}

\newcommand\hS{\hat S}
\newcommand\var{{\rm var}}
\newcommand\cov{{\rm cov}}

\newcommand\cT{\mathcal{T}}

\newcommand\de{\delta}

\newcommand\ta{\tau}

\newcommand\om{\omega}

\newcommand\<{\langle}
\renewcommand\>{\rangle}

\newcommand\ie{\emph{i.e.}}
\newcommand\eg{\emph{e.g.}}

\newcommand\beq{\begin{equation}}
\newcommand\eeq{\end{equation}}
\newcommand\bea{\begin{eqnarray}}
\newcommand\eea{\end{eqnarray}}
\newcommand\bal{\begin{align}}
\newcommand\eal{\end{align}}

\newcommand\fr{\frac}

\newcommand\ap{\approx}

\renewcommand\bal{\mbox{\boldmath$\alpha$}}

\begin{document}

\title{Photon flux and bunching noise from measurement of the shot noise variance}

\author{Richard Lieu$^1$, Michael Stefszky$^2$, and Johann, C. -H. Shi$^3$}

\affil{$^1$Department of Physics, University of Alabama,
Huntsville, AL 35899\\
$^2$Integrated Quantum Optics, University of Paderborn,
Warburger Strasse 100, 33098 Paderborn, Germany.\\
$^3$National Astronomical Observatories, Chinese Academy of Sciences, A20 Datun
Road, Chaoyang District, Beijing 100012, China\\}


\begin{abstract}
We report the experimental observation of photon bunching noise through shot
noise measurements made on a pseudo-thermal state of light using balanced
detection. A full theory describing the measurement is developed, and in
agreement with theory it is found that the shot noise variance in the balanced
signal reproduces the time series of the flux of the primary incoherent beam.
Moreover, when the average power of the pseudo-thermal light is varied, the
balanced detection is seen to track this change. A comparison of direct
detection and balanced detection of the thermal field, shows that the balanced
detection performs at least as well the direct detection and under some
conditions appears to outperform the direct detection.  There is not necessarily a contradiction
with quantum field theory which predicts that at best the performance of the balanced detection
should be equal to the direct detection, because the direct detection process is subject to nonlinearity that has not been excluded by measurements (even though any tests we performed suggest such effects are small).  This is the first time that the bunching noise effect of high
occupation number chaotic light via the shot noise of the field has successfully been
measured, to the point of using it to infer the flux of the field. The findings may be relevant to radio
receiver design, specifically from the viewpoint of sensitivity improvement.
\end{abstract}




\section{Introduction}

In Hanbury-Brown Twiss intensity interferometry \cite{han57}, when a stationary
light beam of high occupation number is divided by a 50:50 beam splitter, the
bunching noise (or classical wave phase noise) patterns in the intensity time
series of the two output beams are identical, because the beam splitter causes
only a relative phase difference of $\pi$ in the amplitudes of the two beams.
This wave aspect of light has extensively been studied over the years, see
\eg~\cite{man95}.  However, due to the indivisibility of photons, the other
noise component, the shot noise \cite{ric16}, in each output beam is not
expected to correlate between the two beams.  Hence a subtraction of the time
series of one output beam from another should yield a zero mean flux, but a
flux
variance that is another measure of the flux of the original (primary) beam at
any given instance.  Recently it has been shown, based on quantum field theory,
that such a method of flux determination can in principle be as accurate as, though not surpassing, direct detection (here-and-after simply referred to as DD); in particular, it can reveal the
bunching noise in the flux of the primary beam just as clearly as direct
detection \cite{zmu14,lie15,nai15}.

The purpose of the work presented here is to experimentally demonstrate, for
the first time, the theoretically expected performance of flux measurement via
the shot noise in the field. The shot noise is measured using a balanced
detection (here-and-after simply referred to as BD) scheme, consisting of a
50:50 beamsplitter followed by the subtraction of the two photocurrents
detected
at each beamsplitter output. In this way we demonstrate a qualitative test for
the predictions of quantum field theory and highlight an alternative method for
radiometry. Whilst balanced detection requires greater experimental complexity
than direct detection, it may come with other advantages such as the ability to
negate large DC and low frequency noise components (such as from electronics)
due to the fact that no DC information is required in the balanced scheme.

It might first be useful to seek a heuristic understanding of key predictions
\cite{zmu14,lie15,nai15}.  Since there are classical bunching noise
fluctuations
in chaotic light, for any sufficiently short interval within which this noise
is
constant, the shot noise mean and variance (as could be measured with a single
detector) should both equal the expected instantaneous photon count rate as
determined by the bunching noise pattern there. In this way the shot noise will
then exhibit a time dependent mean and variance as its amplitude adjusts in
tandem with bunching.  According to \cite{zmu14,lie15,nai15}, this correlation
between the shot noise variance and the classical noise intensity is not
expected to
be removed by the subtraction in the balanced detector, as illustrated in
Figure \ref{cartoon}.  This is why a time series of the shot noise variance, as
measured by the BD setup, is expected to trace the primary flux, including its
bunching noise.

\section{Quantum field theoretic prediction of noise behavior in chaotic light}

We now discuss the specific quantum-field theory predictions.

\subsection{Fully incoherent light}

First is the theoretical expression of the ratio of the
intensity covariance to the square of the mean intensity for stationary light.
If the light is fully incoherent with a Gaussian autocorrelation function of
width $\ta$, eq. (9) of \cite{lie15} indicates that for an intensity dataset
$S_k$ ($J_k$ in \cite{lie15}) directly measured by a radiometer the ratio is
\beq  \fr{\cov(S_k, S_l)}{\<S_k\>^2} =
e^{-t_{kl}^2/\ta^2}; \label{covJ1} \eeq
while for the BD squared difference current,
eq. (32) of \cite{lie15} yielded
\beq \fr{\cov(D_k^2, D_l^2)}{\<D_k^2\>^2} = e^{-t_{kl}^2/\ta^2} + 4\de_{kl}.
\label{covD1} \eeq
In both equations $\cov (A,B) = \<AB\> -
\<A\>\<B\>$ with $\<\cdots\>$ denoting ensemble mean,  $\ta$ is the coherence
time or inverse bandwidth of the incoherent light, and $t_{kl} = (k-l)T$ with
$T$ being one interval of sampling $T < \ta$ but long enough to collect
$\gg 1$ photons per interval.  Thus \beq \hS_k = \fr{1}{T}\int_{(k-1)T}^{kT}
dt\,\hS(t) \eeq with a similar expression for $\hD_k^2$.

\begin{figure*}[!h]
\begin{center}
\includegraphics[angle=0,width=3.8in]{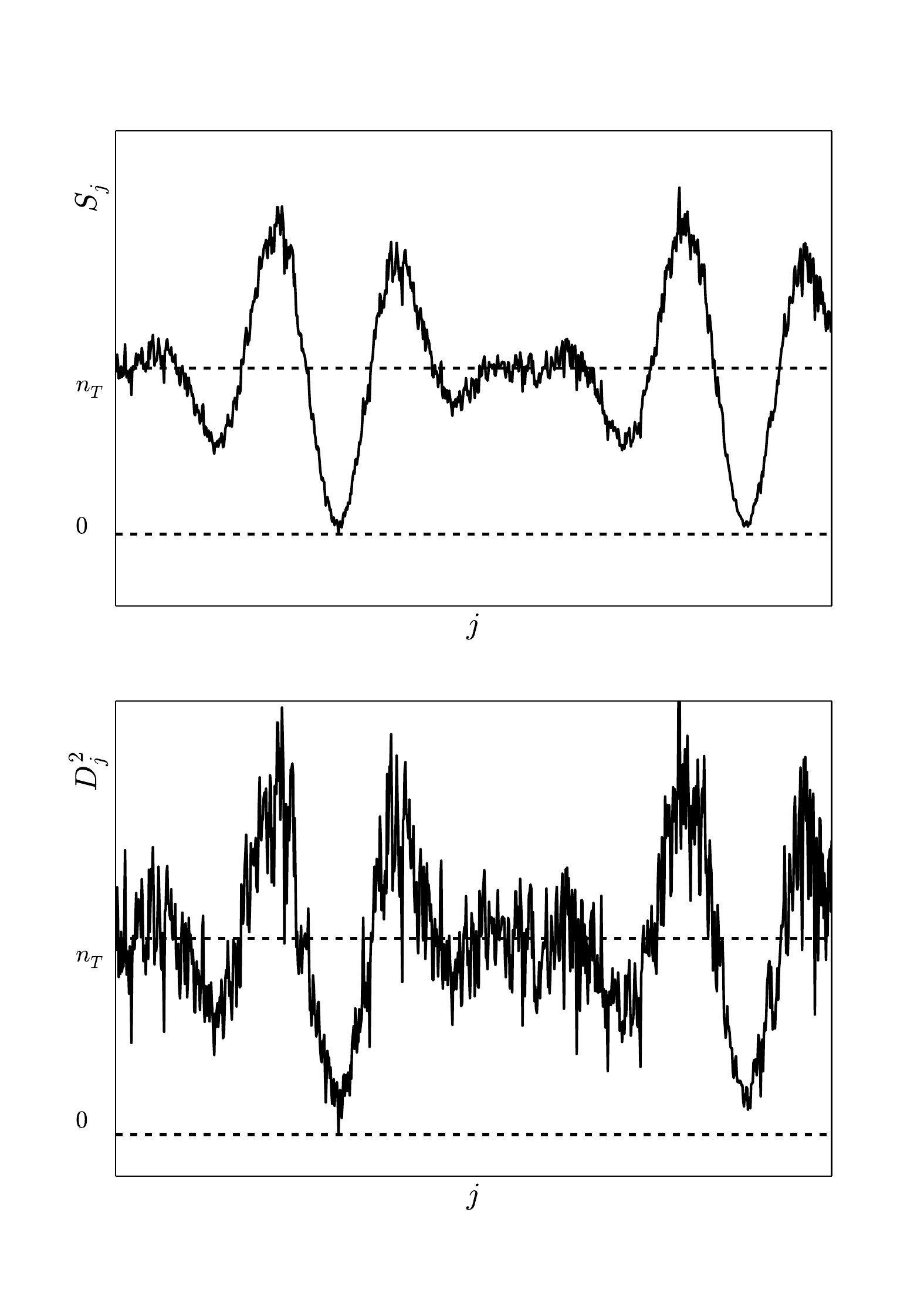}
\end{center}
\caption{Theoretically expected noise characteristics of the intensity of
stationary thermal radiation as measured directly, and via a balanced
(\ie~50:50) beam-splitter detection technique referred to as BD in the text.
The $x$-axis gives the data sample index (hence time, assuming the data are
sampled time contiguously and at equal intervals), and  the $y$-axis the
incident light intensity measured by the two methods: direct is the top graph
and BD the bottom.  Note the BD squared intensity difference $D_j^2$ between
the
two split beams, which equals (in terms of photons counts per bin) to the
variance {\it and} also the ensemble average of the intensity $S_j$ of the
primary incident beam, still exhibits classical bunching noise that correlates
with the direct intensity time series.  Additionally the former is also
expected
to have more shot noise.  This figure illustrates why $D_j^2$ is a measure of
the primary incident flux as well as $S_j$. The BD method can theoretically be
as sensitive as the direct, but cannot surpass it.}
\label{cartoon}
\end{figure*}

To interpret (\ref{covJ1}) and (\ref{covD1}), one may revisit Figure
\ref{cartoon}.  In the top schematic, the direct time series, the rapid
fluctuations of the shot noise is seen to be dominated by the slower bunching
noise variation, the characteristic width of the latter is $\ta$ and is given
by
one of the broad peaks shown (the insignificance of the shot noise is due to
the
assumption that the light has high photon occupation number, \ie~the number of
photons per interval $\ta$ is $\gg 1$).  Thus the autocorrelation function of
$S_k$ has as its chief feature a Gaussian of width $\ta$.  On the other hand,
the bottom schematic of Figure \ref{cartoon} shows that the BD time series has
shot noise and bunching noise of approximately the same strength; the former
has
a sharp autocorrelation function spanning one bin width $T$, while the latter
has width $\ta$ which spans many intervals $T$ but height equal to the former.

Moreover, (\ref{covJ1}) shows that for direct radiometry the variance obtained
by setting both $k$ and $l$ to the same index $j$ relates to the mean squared
intensity as \beq \fr{{\rm var} \<S_j\>}{\<S_j \>^2} = 1; \label{varS1} \eeq
while the same for (\ref{covD1}) shows that for a BD measurement \beq \fr{{\rm
var} \<D_j^2\>}{\<D_j^2\>^2} = 5. \label{varD1} \eeq  In the extreme opposite
limit of flux estimation, by averaging over a large interval $\cT = NT$
comprising $N \gg 1$ intervals of $T$, (\ref{covJ1}) gives
\beq \fr{\var (\overline{S_\cT})}{\<\hS_j\>^2} \ap
\fr{\sqrt{\pi}\ta}{\cT}=\fr{\sqrt{\pi}\ta}{NT};
 \label{errordirect1}\eeq
while (\ref{covD1}) gives
 \beq \fr{\var(\overline{D_\cT^2})}{\<\hD_k^2\>^2} =
\fr{1}{N}\left(\fr{\sqrt{\pi}\ta}{T}
+4\right), \label{varDD}\eeq
where
\beq \overline{S_\cT} = \fr{1}{N}\sum_{k=1}^N S_k \eeq and
\beq \var(\overline{S_\cT}) = \fr{1}{N^2}\sum_{k,l=1}^{N} \cov(S_k,S_l);
\eeq and similar expressions for $\overline{D_\cT^2}$.

Now (\ref{errordirect1}) is the radiometer equation
for the ultimate sensitivity limit of the direct flux detection of fully
incoherent light of high occupation number, a limit set by the bunching noise
in
the direct time series.  The underlying physics is simply that bunching noise
variances add on timescales greater than the coherence time $\ta$, when the
different segments of the time series have become independent (as seen in the
top half Figure \ref{cartoon}).  Evidently, (\ref{varDD}) indicates that on
sufficiently long timescales $\cT$ the sensitivity of the BD measurement method
can approach the direct without surpassing it, because the BD time series is
still expected to carry bunching noise of the same amplitude, as discussed in
the previous paragraph.  The situation is essentially the same for partially
incoherent light, which we now turn to.

\subsection{Partially incoherent light}

If the light is partially incoherent, as is the case of the experiment we
performed, then by the derivation shown in the Appendix (which contains the
proof of all key formulae in this subsection) the ratios of (\ref{covJ1}) and
(\ref{covD1}) would become (ignoring the shot noise contribution which is
negligible because the beam has high occupation number)
\beq \fr{\cov(S_k,
S_l)}{\<S_k\>^2} = \nu^2 e^{-t_{kl}^2/\ta^2},  \label{covJ} \eeq
and
\beq \fr{\cov(D_k^2, D_l^2)}{\<D_k^2\>^2} = \nu^2
e^{-t_{kl}^2/\ta^2} + 2(1+\nu^2) \de_{kl} \label{covD}\eeq
respectively, where
$\nu \leq 1$ is the relative amplitude of incoherent fluctuations defined as
\beq \nu = \fr{\sqrt{{\rm var} \<S_j\>}}{\<S_j\>}. \label{nuu} \eeq
As will soon be discussed, the light we employed for our tests have $\nu < 1$
and $\nu^2 \ll 1$.  Thus the covariance of the direct and BD time series $S_k$
and $D_k^2$ should, by (\ref{covJ}) and (\ref{covD}), respectively be a
Gaussian
of width $\ta$ and height $\nu^2$, and the same Gaussian with a narrow central
spike of of height $2(1+\nu^2)$ and spanning the width of one interval $T$.

The corresponding variances (\ref{varS1}) and (\ref{varD1}) are still evaluated
by the same procedure as there (see Appendix), but now assume the (more
general)
expressions
\beq \fr{{\rm var} \<S_j\>}{\<S_j \>^2} = \nu^2 \label{varS} \eeq
and this is the radiometer equation (direct detection sensitivity) of partially
incoherent light, and
\beq \fr{{\rm var} \<D_j^2\>}{\<D_j^2\>^2} = 2+3\nu^2;
\label{varD} \eeq
while the long term variances now lead to the following form
for their ensuing noise-to-signal ratios:
\beq \fr{\var (\overline{S_\cT})}{\<\hS_j\>^2} \ap
\fr{\nu^2\sqrt{\pi}\ta}{\cT}=\fr{\nu^2\sqrt{\pi}\ta}{NT},
 \label{errordirect}\eeq
and
\beq \fr{\var(\overline{D_\cT^2})}{\<\hD_k^2\>^2} =
\fr{\nu^2\sqrt{\pi}\ta}{NT}
+\fr{2(1+\nu^2)}{N}. \label{varDD1}\eeq
Beware when $\nu^2 \ll 1$ {\it but} $N \gg 1$ is not too large, the
$2(1+\nu^2)/N$ term of (\ref{varDD1}) that depicts the BD shot noise variance
exceeds the bunching noise variance of the preceding term, despite the
inequality $T \ll \ta$ which {\it always} applies to our sampling strategy.
For
{\it very} large $N$, the sensitivity of the BD method is governed by the first
term on the right side of (\ref{varDD1}), as the last two terms become
sub-dominant.  Since the first term originated from the Gaussian
autocorrelation
of the BD time series, {\it viz.}~the $\nu^2 e^{-t_{kl}^2/\ta^2}$ term of
(\ref{covD}), this indicates the BD signal is just as much contaminated by
bunching noise as the direct (as (\ref{covJ}) also has this term).  Observe
that
when $\nu^2 \ll 1$, if $N$ is not sufficiently large the $2(1+\nu^2)/N$ term of
(\ref{varDD1}), depicting the BD shot noise variance, can exceed the BD
bunching
noise variance of the first term; this feature does not apply to fully
incoherent light, where $\nu = 1$ and (\ref{varDD}) reduces to (\ref{varDD1}).
Ultimately it is the inability of the BD technique to remove or reduce the
bunching noise in the original incident beam, that prevents one from employing
this method to surpass the sensitivity limit of the radiometer equation.

In the case of partially incoherent light, as we shall see shortly, the
bunching noise in both time series remain equal to each other in amplitude,
while the shot noise exceeds the bunching noise in the BD time series even
though it remains subdominant in the direct series.

\section{The experiment}

In the previous section we presented the status of recent quantum optics
calculations as an extension of earlier semi-classical treatment of $n$-point
correlation functions in which the amplitudes are classical $c$-numbers
(\eg~\cite{wan89}), and summarized their specific predictions of flux
measurement sensitivity comparison between two methods.
Despite the theoretical consensus among the 3 most recent papers on the
subject:
\cite{zmu14,lie15,nai15}, there has not been any deliberate experimental
verification of their predictions.

Previous experiments have, however, used similar experimental setups to
characterize states of light. For example, the standard setup used for measuring the shot noise level in homodyne detection (in which the vacuum is chosen as the signal field and is mixed with a bright coherent state, known as the local oscillator) is identical to balanced detection of a coherent state. Homodyne detection has been used for many years to perform optical tomography
\cite{lvo02,smi93} and has even been used previously to measure the
photon-bunching characteristics of a given field \cite{gro07}. Furthermore, a similar scheme has been used previously to produce random numbers through the measurement of vacuum fluctuations \cite{she10}. In this experiment, the Fourier spectrum of the subtracted
intensity was seen to have an autocorrelation function comprising a narrow
central spike above a zero background, as one would expect from a coherent
local
oscillator. In contrast to these experiments, the method presented here
involved the mixing between a thermal state of light with the vacuum field entering the
empty port of the beamsplitter. Furthermore, the detection schemes vary
significantly in the various experiments according the the information that one
wishes to obtain.

In our experiment a direct comparison between the direct detection method and
balanced detection for determining the flux of a given field is made. In order
to provide a more meaningful comparison between the two methods, a stable
reference field is also measured. Through comparison of the signals measured
via
balanced and direct detection with the measurements from the reference field,
the performance of the two detection schemes is directly compared.

\subsection{Basic setup and measurement strategy}

\begin{figure*}[!h] 
  \centering
    \includegraphics[width=1\textwidth]{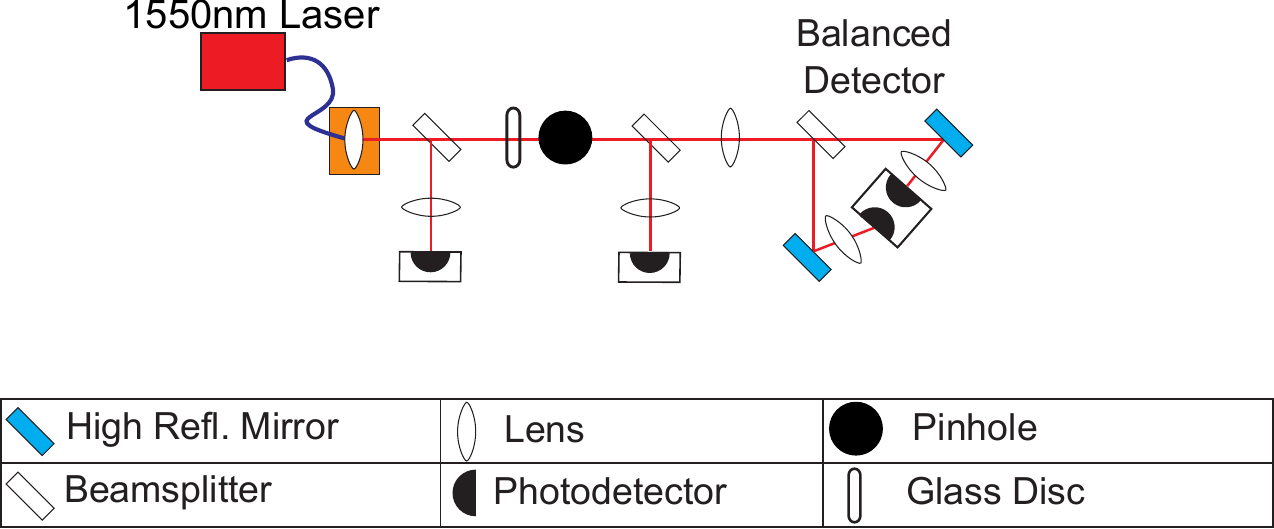}
\caption{Experimental layout for the near simultaneous flux measurements of a
beam of coherent light .  The ground glass is rotated off (optical) axis to
produce pseudo-thermal, or partially incoherent, light with classical bunching
noise extending to $\sim 1-2$ MHz via scattering.  The photodetectors are described in more detail in the text.  In such a setup, the directly measured flux of the laser light serves as the reference flux; against which the flux of the same light after it
loses coherence (as a result of passage through the ground glass), as measured
by the direct and BD detection methods, are compared.}
 \label{glass}
\end{figure*}

An overview of the experimental layout is shown in Figure \ref{glass}. The
laser source is a low noise (RIO Grande) laser. An initial tap-off of a few percent from this
field provided the \textit{reference detection}. Any changes in average field
power can be detected from this tap-off. Also note that the changes measured
here will be proportional to any average field changes in the following
detection schemes.

\begin{figure*}
\begin{center}
\vspace{-1cm}
\includegraphics[angle=0,width=8in]{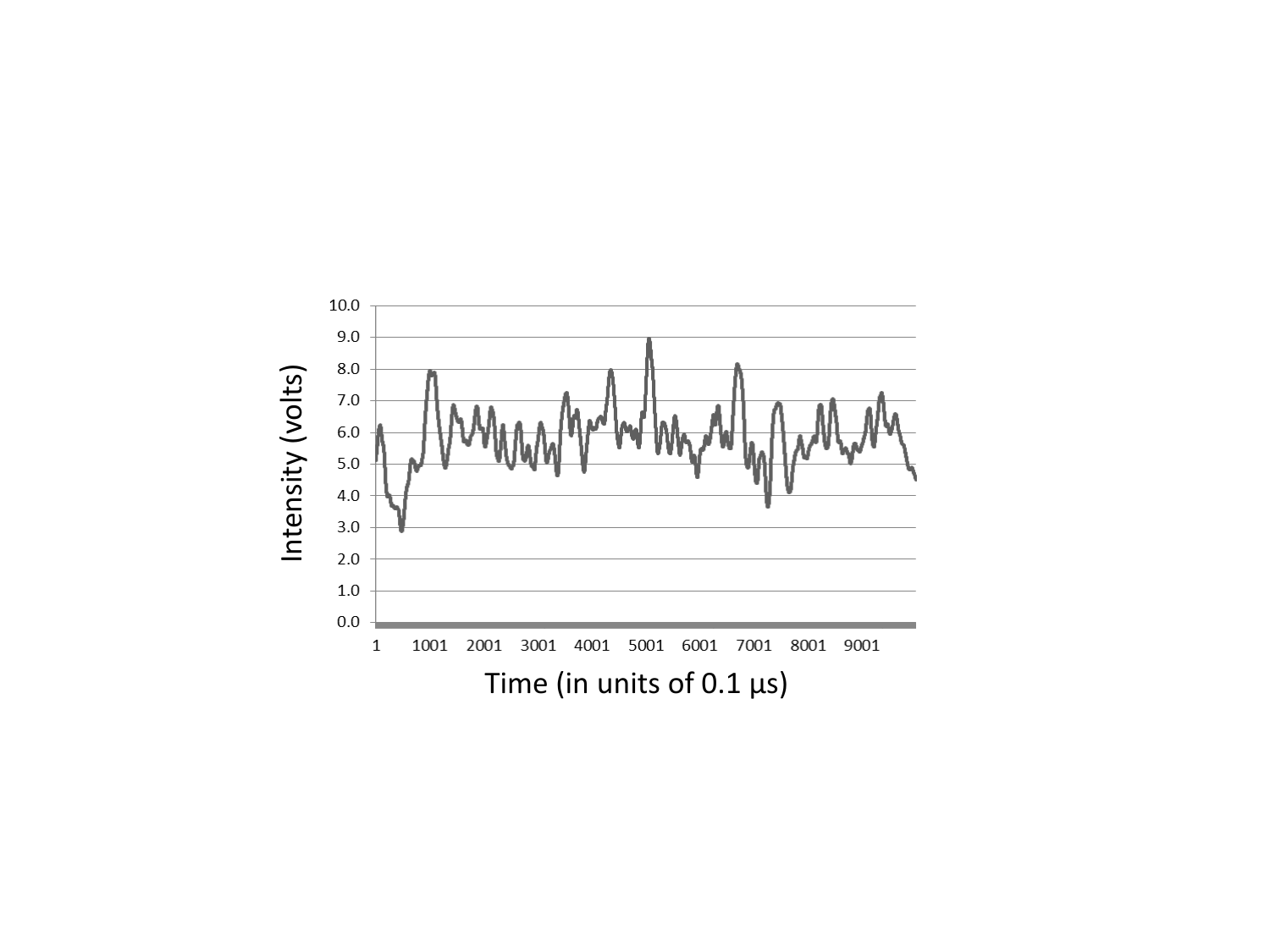}
\vspace{-3cm}
\end{center}
\caption{The time series of partially incoherent light as detected by the direct
detection scheme, with the classical bunching noise generated by a rotating
ground glass plate, see Figure \ref{glass}.  The light is partially incoherent
because the ratio $\nu$ of the standard deviation to the mean intensity (see
(\ref{nuu}) and (\ref{nu})) is less than unity.  This type of light is used in
our experimental tests of the sensitivity of direct versus BD flux measurement
technique. }
\label{series}
\end{figure*}

\begin{figure*}[!h]
\begin{center}
    \includegraphics[angle=0,width=6in]{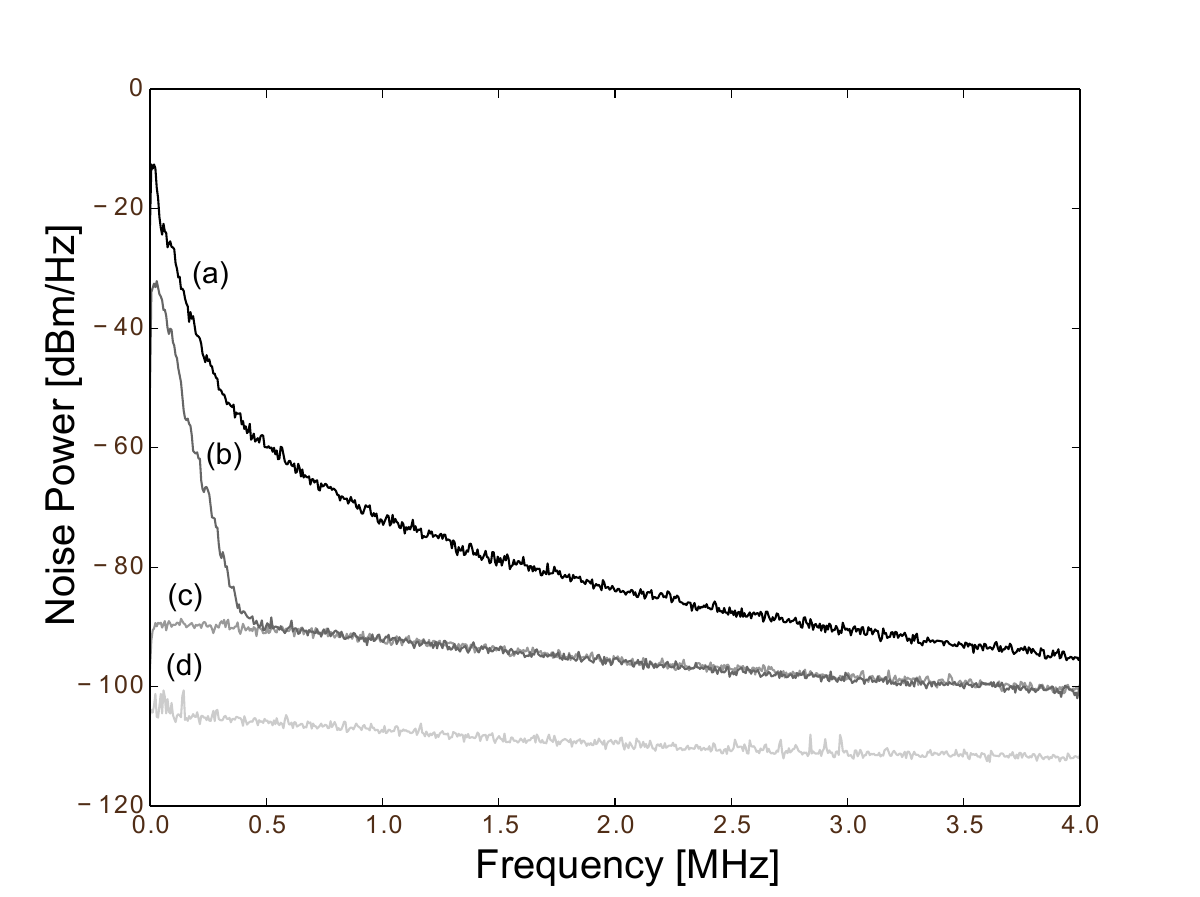}
\end{center}
 \caption{Spectra of the BD detector output voltage with different fields
 incident on the diodes showing typical characteristics of the fields. Trace
(a) is the noise spectrum of the pseudo-thermal light (passing through the
rotating plate), when all the incident power is on a single diode. Trace (b) is
the pseudo-thermal light noise spectrum when the optical power is distributed
equally between both diodes and BD subtraction is at a maximum. Trace (c) is
when the rotating plate is removed and the power is equal on both diodes such
that the shot noise of the coherent field is measured. Trace (d) is the dark
noise of the detection system, whereby no light is incident on the detectors.
The incident power for traces (a),(b), and (c) are equal.  Note how both (b)
and
(c) merge at the shot noise level when frequencies are high.  All traces were
taken on a spectrum analyzer with an RBW of $3$ kHz and VBW of $10$ Hz.  In
order to show the dark noise clearance no gain correction was applied to the
traces, which is why the shot noise asymptote is not flat. This figure shows
how
the shot noise level reached (at high frequencies) by the BD flux measurement
method stands sufficiently above the dark noise (DN clearance is a minimum of 8dB at 20MHz)}.
    \label{Disc}
\end{figure*}

After this inital tap-off the field passes through a rotating glass disc and a
pinhole. This setup is used to produce pseudo-thermal light following the
method
of \cite{mar64}, with a photon bunching bandwidth of approximately 0.2 MHz. Note that the use of pseudo-thermal light is necessitated by the limitations of the experimental setup, as it is necessary to find a condition where the power in the field is high enough that the shot noise component is measurable, but also where the thermal noise component is not so large that it cannot be subtracted in the frequency band of interest. The effect of the plate on the time and frequency domains is shown in Figures
\ref{series} and \ref{Disc} respectively. The state of light so produced has a
relative amplitude of incoherent fluctuations
\beq \nu  \ap 0.15 \label{nu} \eeq
where $\nu$ was defined in (\ref{nuu}). The properties of the pseudo-thermal
light are chosen such that the limited subtraction of the balanced detector is
sufficient to remove all of the classical noise in the frequency band of
interest whilst ensuring that the average power through the pinhole is large
enough such that the balanced detector can measure the shot noise level above
the noise floor of the detector.

After the light is thermalised, another beamsplitter takes a tap-off of the
pseudo-thermal light. This light is directed to a single detector where a
\textit{direct detection} is made. The light passing through the beamsplitter
simultaneously undergoes a \textit{balanced detection}. The balanced detector
is
used to subtract the classical (or photon bunching) noise and allow shot noise
level measurements. A direct current subtraction method is used to maximise the
subtraction afforded by the balanced detector  \cite{ste12}. The output of the
balanced detector then undergoes high-pass filtering (at 1 MHz) in order to
remove residual low-frequency noise sources, such as beam jitter \cite{ste12}.
This setup allows us to make simulataneous comparisons between the average
power
in the (nearly) coherent reference field and the direct and balanced detection
schemes.

\begin{figure*}[!h] 
  \centering
    \includegraphics[angle=0,width=8in]{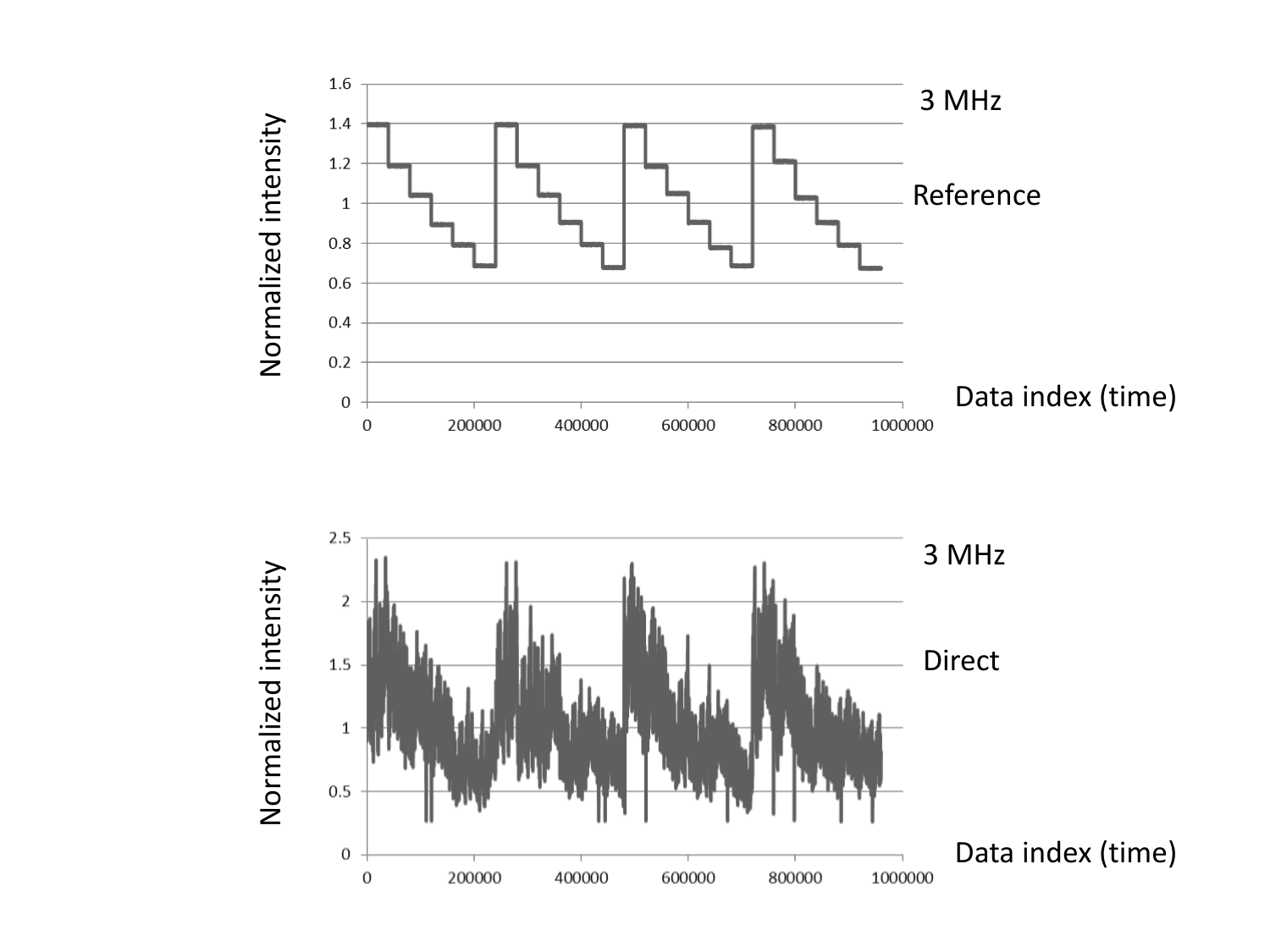}
\caption{The entire (concatenated) data run with a sampling interval of 0.167
$\mu$s, as measured via the reference detector (top) and direct detector after
passing through the rotating disc (bottom). The data from the reference
detector
shows how the laser power varies over time, and the data from the direct
detector (bottom) shows the effect of photon-bunching contamination due to the
rotating disc. All intensities shown are normalized to a mean of unity over the
entire time series.}
    \label{pedestal}
\end{figure*}

The goal of the experiment is to track power changes of an incident laser field
using both direct and balanced detection schemes simultaneously as illustrated
in Figure \ref{glass}. We are interested in the average flux of the field,
rather than the fluctuations caused by the thermal nature of the light, and to
this end the power exiting the laser is varied via control of the laser
amplifier gain. In this way, the average power of the laser is halved from some
initial value in 6 steps. The detected reference field provides an accurate
measure of this change in power, as shown in Figure \ref{pedestal}. At each of
the six power levels 40,000 data points are recorded, resulting in a total of
240,000 data points. This entire measurement of 240,000 points is repeated at
least two more times. For the analysis all of this data is concatenated (See
Figure \ref{pedestal}). This entire measurement procedure is completed with
three different sampling intervals (of 0.167, 0.1 and 0.025 $\mu$s), such that
the maximum frequency of the Fourier transforms of these frequencies are 3, 5
and 20 MHz respectively.


\begin{figure*}[!h]
  \centering
    \includegraphics[angle=0,width=7in]{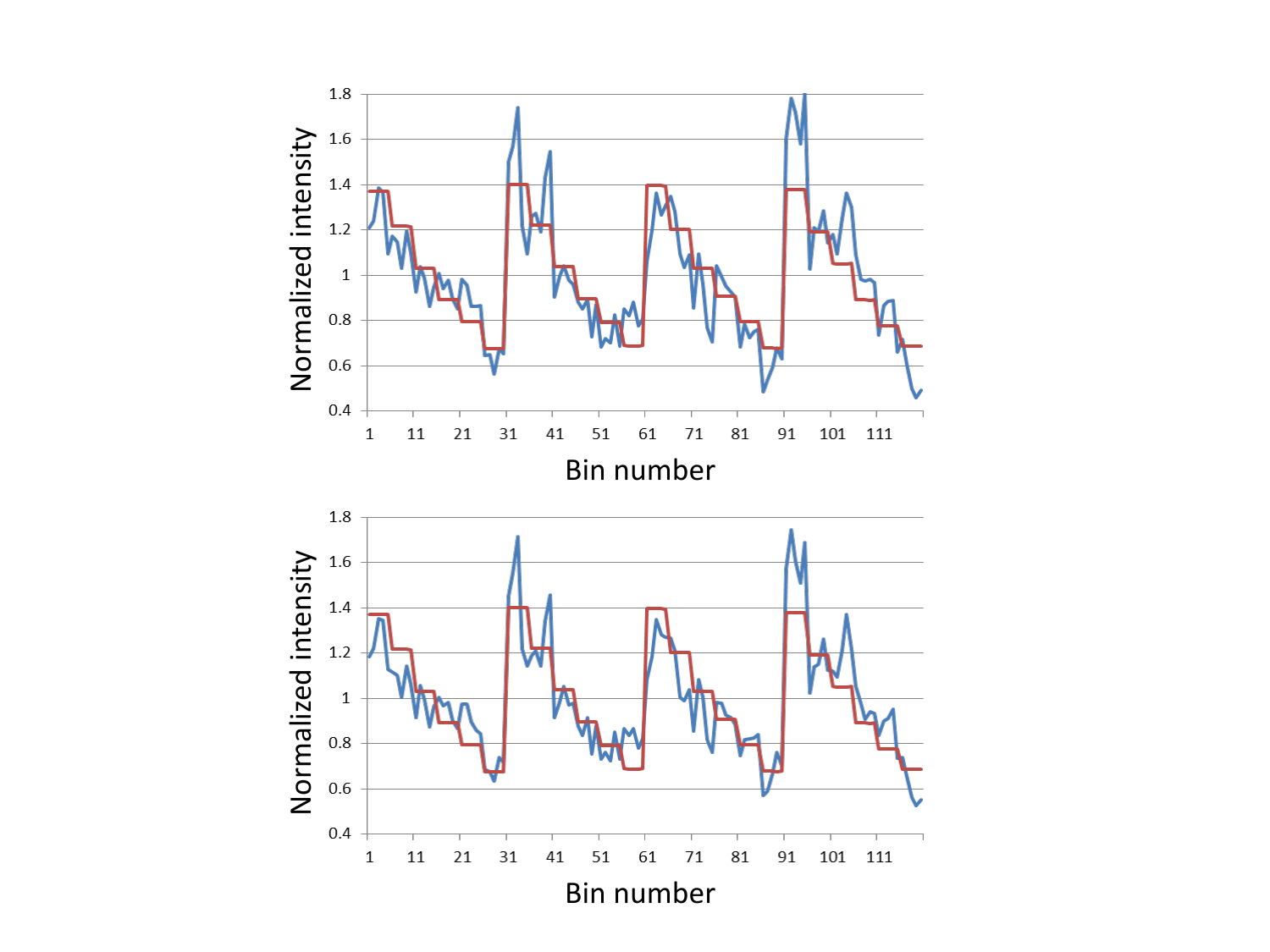}
 \caption{
The direct (top blue) and BD (bottom blue) time series with sampling interval
of
0.025 $\mu$s, and bin-averaged every 800 (\ie~$N=800$) of the 960,000 samples.
The BD time series is obtained after the Fourier transform, filtering and then
inverse Fourier transform.
The two series are to be compared to the underlying directly measured coherent
reference signal (red).
Note that each small wiggle superimposed upon the underlying pedestal is due to
bunching noise,
as the largeness of $N$ ensured that the shot noise component in the BD is
ironed out (this component is negligible in the direct series irrespective of
the size of $N$).
The r.m.s. percentage difference between the BD flux and
reference flux is 1.57 \% over the entire time series, while the same between
direct and reference is $> 50$ \% higher, {\it viz.} 2.37 \%.
}
    \label{pedestal800}
\end{figure*}

\subsection{Photodetectors}

All detectors used in the setup consist of a Hamamatsu G12180-003 (or two in the case of the BD) photodiode followed by a transimpedance gain stage and an output gain stage, both utilising AD829 opamps with a 15V supply. The output voltage of all devices is kept below \textbf{10V} to stay within the linear range of the opamp. All optical powers are also kept below the 6mW linear range limit of the photodiodes. The reference field detector has a 20k opamp gain stage, resulting in a bandwidth of around 6MHz and the maximum incident light power is approximately 200$\mu$W, while the direct detector has a 2k gain stage, giving a bandwidth of around 10 MHz, with a maximum incident power of 2mW (note that the power is varied in the measurements). The direct and reference detection are limited by the oscilloscope noise (which is approximately 3 orders below the signal level) as the dark noise levels of these detectors is below the electronic noise of the oscilloscope. The BD detector has two photodiodes in a current subtraction scheme (described in \cite{ste12}) followed by a 20k transimpedance stage and a 1.2 MHz high pass filter before the output gain stage to remove residual noise at low frequencies. This scheme provides very high subtraction of up to 60 dB and a bandwidth of 6MHz, but does not allow for measurements of the photocurrents from the individual photodiodes.

\begin{figure*}[!h]
	\centering
	\includegraphics[angle=0,width=7in]{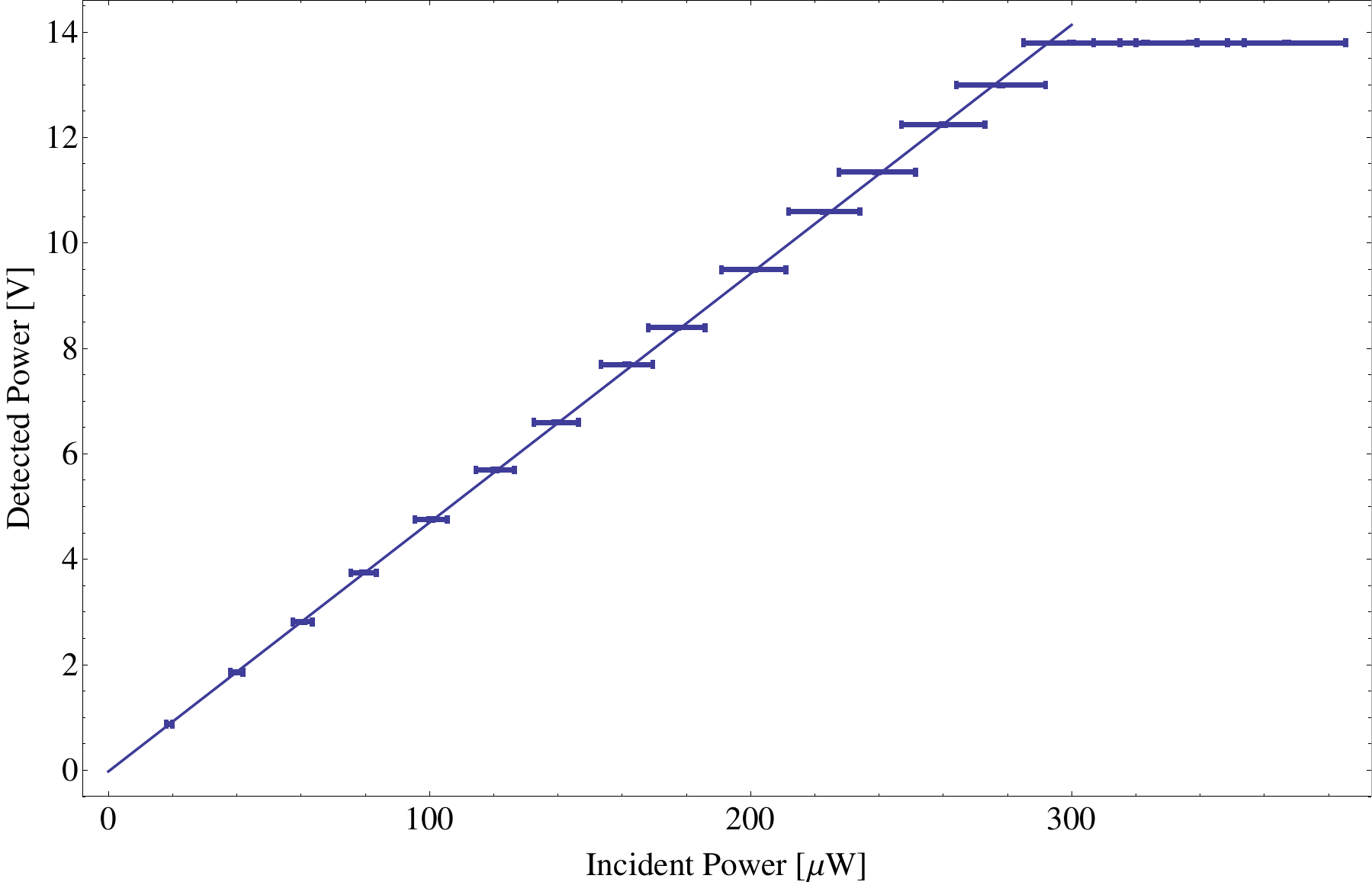}
	\caption{Measured linearity of the reference detector. The power before incidence (measured using a Thorlabs S132C detector with 5\% uncertainty) is compared to the power measured converting the photodetector voltage.}
	\label{detcal}
\end{figure*}

All measurements are taken simultaneously using the same oscilloscope and the appropriate internal low pass filter is implemented to limit aliasing for all measurements. The measurements from both BD and DD detection schemes are compared to the reference detector in order to judge their performance. In this way it is possible to remove absolute errors that would otherwise be introduced, due to the fact that we need only measure the relative changes in flux. It is then of paramount importance that this detector is reliably operating in the linear regime, as all other measurements are referenced to the performance of this detector. As stated previously, to ensure that this is the case, no more than approximately 200$\mu$W is incident on the detector, and the output voltage under operation is limited to 10V. This ensures that both the photodiode and opamps are well within their linear operating range of 6mW and $\approx$13V respectively. The measured linearity of this detector is illustrated in Figure \ref{detcal}. The linear fit to the data has a gradient of $4.7 \times 10^{-2}$ V/$\mu$W with a standard error of $1.5 \times 10^{-4}$ V/$\mu$W and a p-value of $1.8 \times 10^{-26}$. It is evident that the detector is in the linear regime below around 300 $\mu$W of input power.

The slew rate requirements of the detectors are strictest on the DD scheme because the measured voltages are highest and the signal is strongly oscillating. In order to ensure that this detector was not slew rate limited it is simulated using TINA-TI SPICE. These results show that the expected slew rate $SR$ of the DD detector is approximately 1V/70ns. From this value the standard slew rate definition can be used to determine the maximum amplitude $V_{pk}$ of a sinusoid with frequency of $f=20$MHz that can be reliably measured, which is found to be $SR/(2 \pi f) \approx 115$mV. This is approximately one order of magnitude higher than the voltage changes seen in the detector on these timescales (50 ns).

Finally, the use of three detection schemes with varying gains, incident optical powers and designs, provides additional confidence in the performance of the detectors. Owing to the very different operational conditions of the three systems, measurements in which the results from two or more detectors coincide virtually guarantees that these detectors are operating in the nonlinear regime. This is because it is virtually impossible for two very different nonlinear systems to track one another over a wide range of inputs.

\section{Bunching noise from shot noise measurement}

We now show in Figure
\ref{pedestal800} the BD and direct time series when the data is binned to bin
sizes of $N=800$ in order to smooth out the variance of the variance of the
measured shot noise, \ie~to suppress the effect of $2(1+\nu^2)/N \ap 2/N$ term
of (\ref{varDD1}) by letting $N$ become large.  In this way, each of the
relatively high frequency fluctuations that were not part of the pedestal
pattern of the laser light, Figure \ref{pedestal}a, is due {\it solely} to
photon bunching noise (note that there is essentially no shot noise in the
direct time series because the incoherent light has very high occupation
number).  It is then obvious that that the bunching noise as revealed by the
direct and BD methods correlate very well, practically there is one-to-one
correspondence between every peak and trough of fluctuation.  It can also be
seen, surprisingly, that the BD method seems to involve less bunching
fluctuations.  In fact, a least square test revealed that Figure
\ref{pedestal800}b has a smaller $\chi^2$ difference from Figure \ref{pedestal}a
than Figure \ref{pedestal800}a by $\ap 25$ \%.  This difference could be for the reasons outlined in section 3.2, where it was demonstrated that nonlinearity in the direct detection process is a possible cause we have not succeeded in ruling out, even though the effect is unlikely to be large enough to account for the observed anomaly.

\section{Correlation analysis}
\begin{figure*}[!h] 
  \centering
    \includegraphics[angle=0,width=5in]{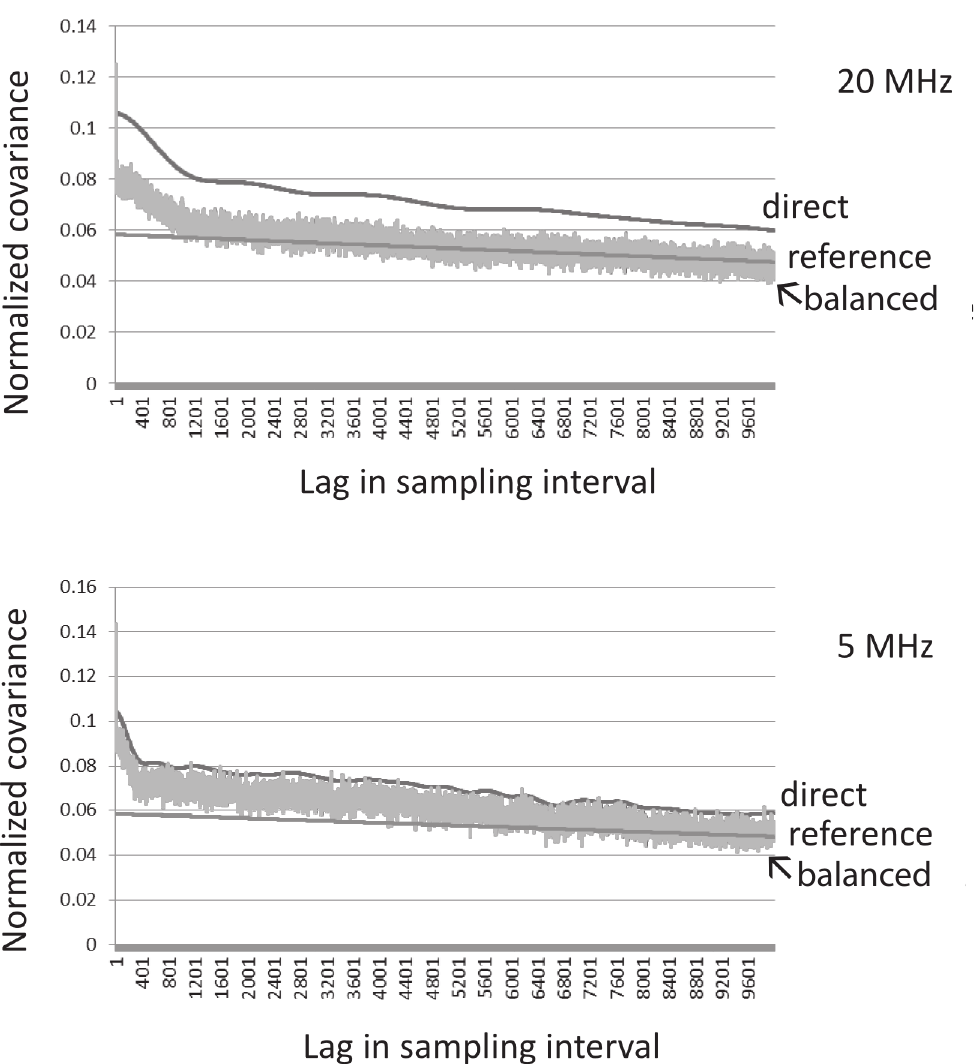}
 \caption{The correlation measurements of the direct,
 BD squared (lightest color), and reference signals for two sampling
frequencies.  The normalized covariance is defined as the covariance divided by
the product of the means,
\ie~$\cov(A,B)/\<A\>\<B\>$
where
$\cov (A,B) = \<AB\> - \<A\>\<B\>$
and the ensemble mean $\<\cdots\>$ is estimated by the sample mean.  The central
two lag intervals are omitted from all data to avoid the very tall shot noise
peak in the BD squared signal, {\it viz.} the $2\de_{kl}$ term of (\ref{covD})
(there is some spill-over into the next several bins, resulting in the much
shorter spike on the extreme left).  Both the direct and BD correlation
functions have a long tail above the reference level, indicating that the width
$\ta$ of the central Gaussian-like peak is actually the minimum coherence time,
\ie~there are other (larger) coherence scales present in the bunching noise.
The suppression of the correlation power of the BD squared signal relative to
the direct signal is evident.  It indicates the BD process reduces bunching
noise, more so at higher sampling frequencies, contrary to theoretical
predictions.}
    \label{PedestalACFs}
\end{figure*}

\begin{figure*}[!h]
  \centering
    \includegraphics[angle=0,width=7in]{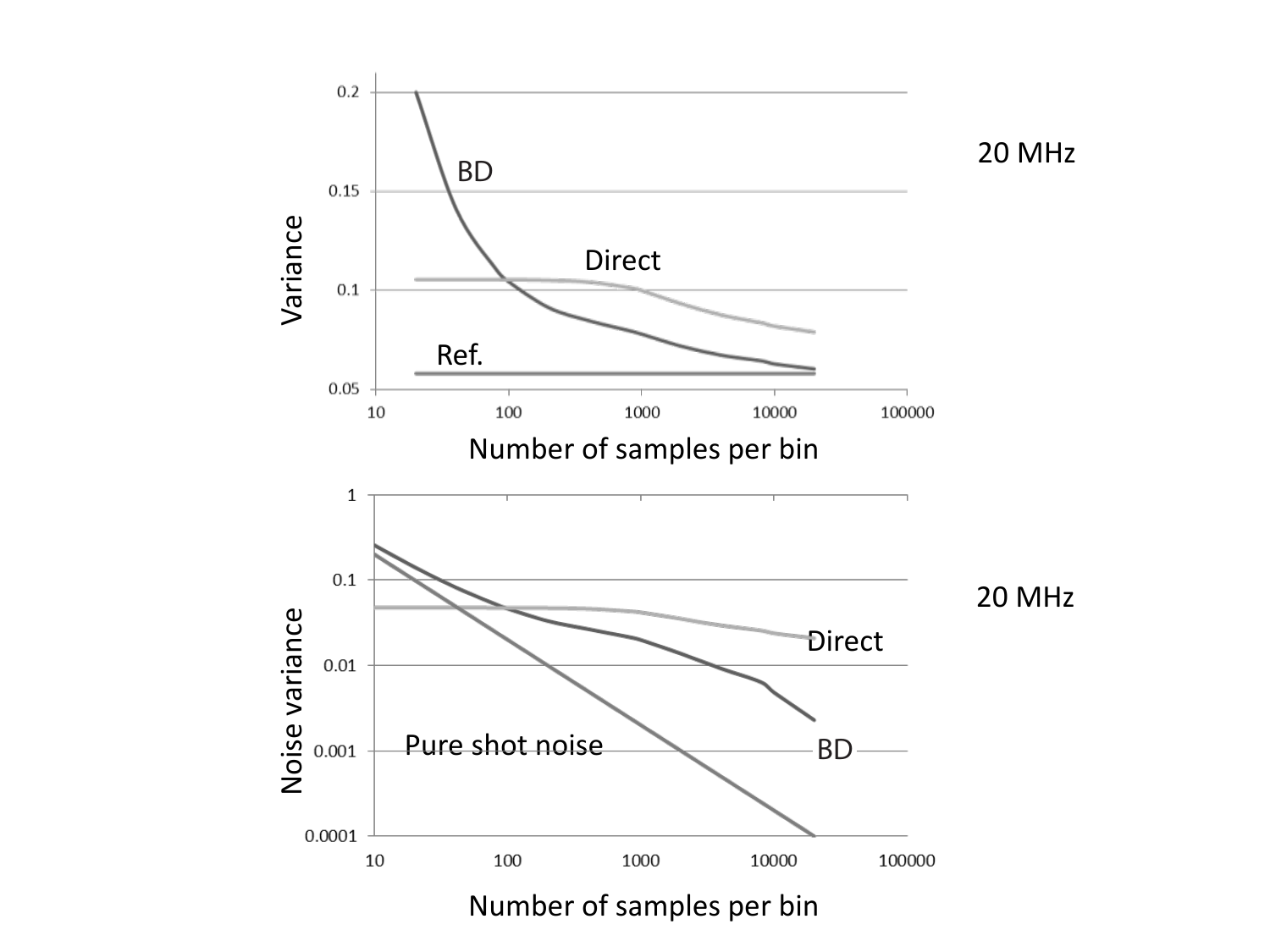}
\caption{Variance in the direct time series of the coherent reference source
and the direct and BD detected time series of the same after the light
partially lost coherence (all variances are defined by the left side of
(\ref{errordirect}) and (\ref{varDD1})), each series is sampled at the interval
of 0.025 $\mu$s (resulting in a maximum Fourier frequency of 20 MHz), and
bin-averaged into various large time intervals given by the $x$-axis.  Each
variance is normalized to the mean-squared flux for meaningful comparison.
Note that the excess variances above the laser reference value are due to
bunching noise only in the direct, and shot noise and bunching noise in the BD
(with latter assuming prominence only towards smaller number of samples per
bin,
$N$), signals.
This excess noise variance is plotted in the lower graph.}
    \label{var20MHz1}
\end{figure*}

\begin{figure*}[!h]
  \centering
    \includegraphics[angle=0,width=7in]{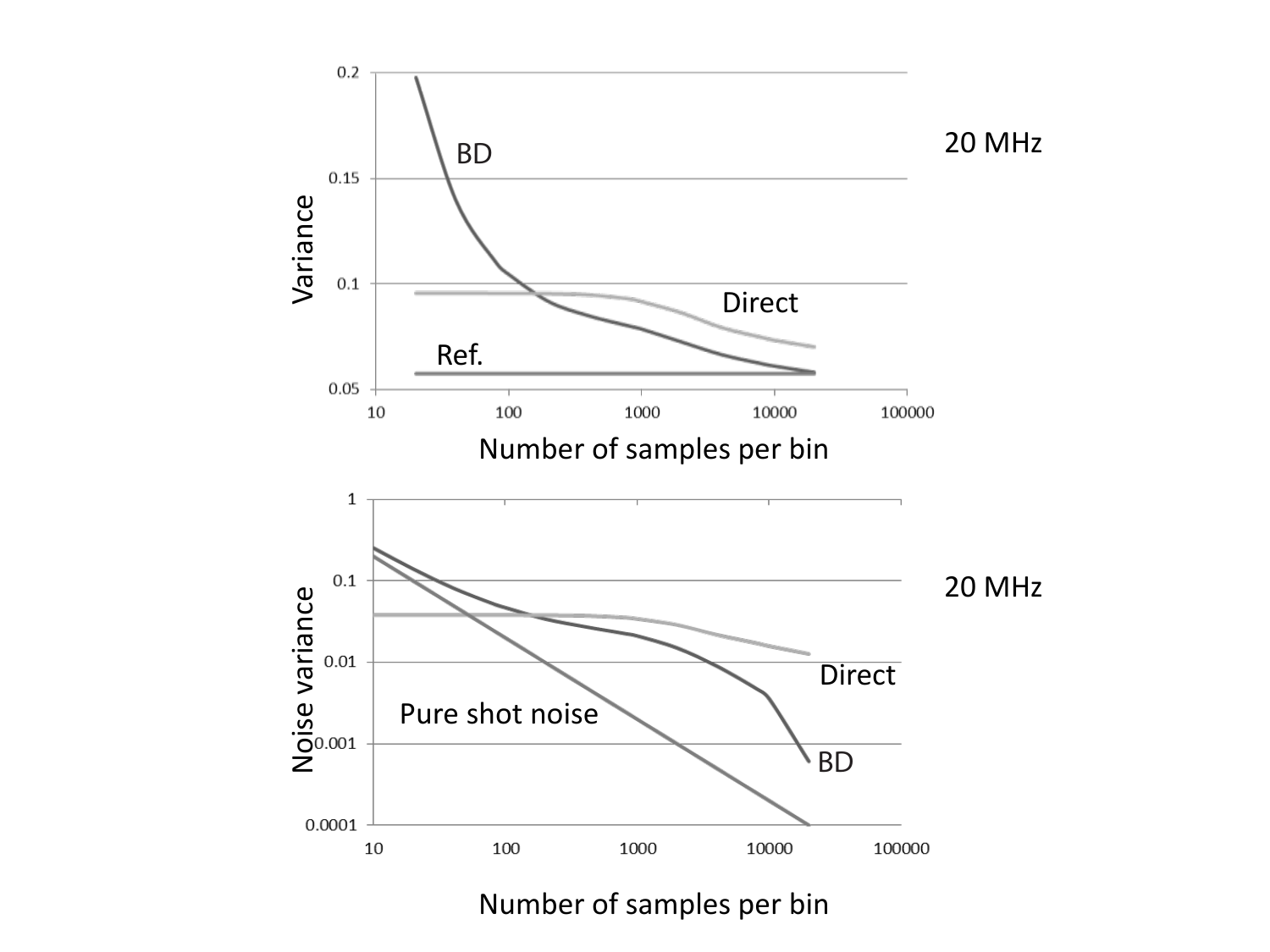}
 \caption{As in Figure \ref{var20MHz1}, except for an independent data set of
 720,000 samples, also acquired with a sampling interval of 0.025 $\mu$s.}
    \label{var20MHz2}
\end{figure*}

\begin{figure*}[!h]
  \centering
    \includegraphics[angle=0,width=7in]{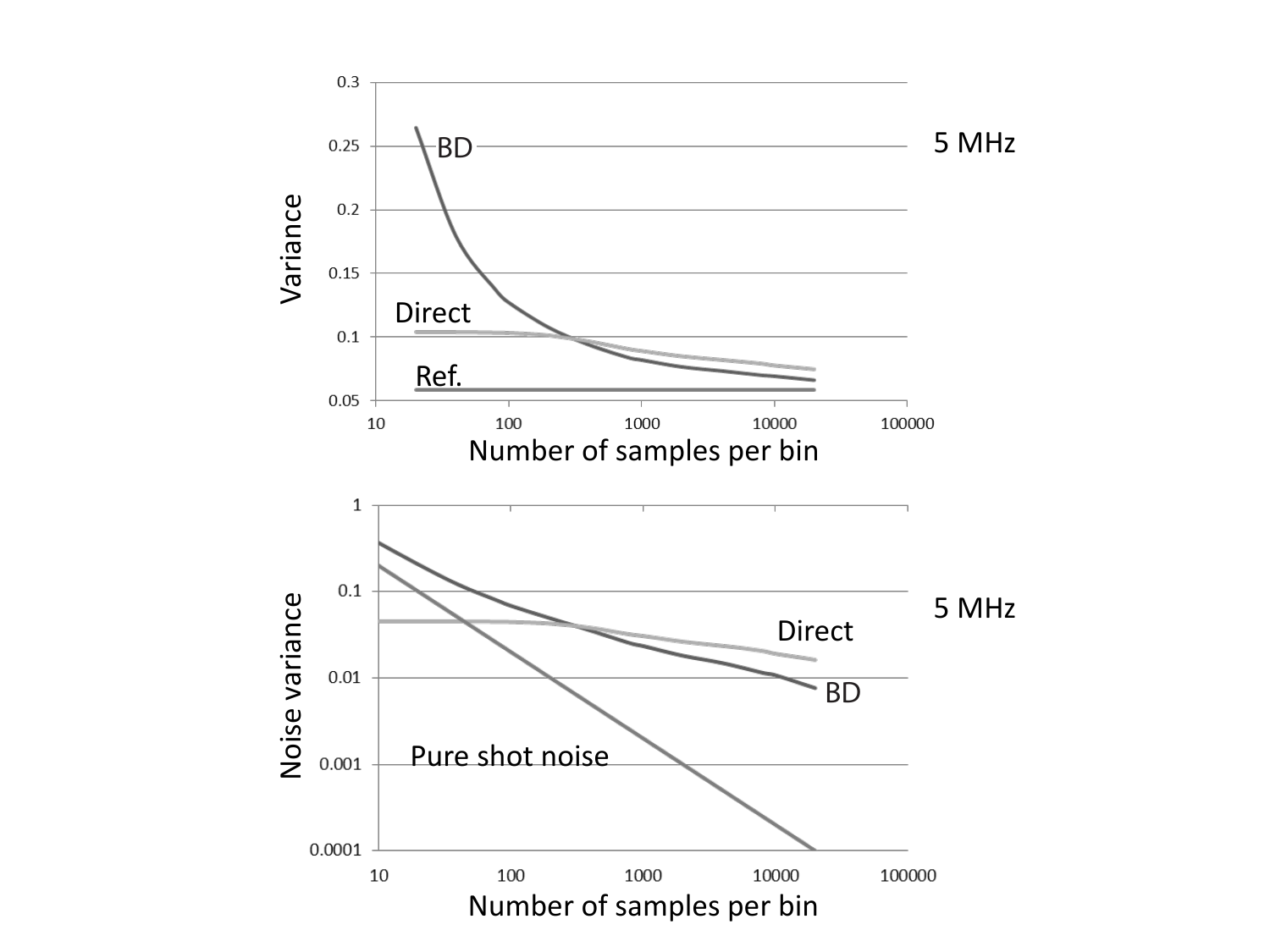}
 \caption{As in Figure \ref{var20MHz1}, except for an independent data set of
 720,000 samples acquired at the sampling interval of 0.1 $\mu$s (maximum
Fourier frequency of 5 MHz).  Note how the reduction in the BD variance w.r.t.
the direct for large bin sizes is not as marked as in the 20 MHz sampling case
of Figures \ref{var20MHz1} and \ref{var20MHz2}. }
    \label{var5MHz}
\end{figure*}

\begin{figure*}[!h]
  \centering
    \includegraphics[angle=0,width=7in]{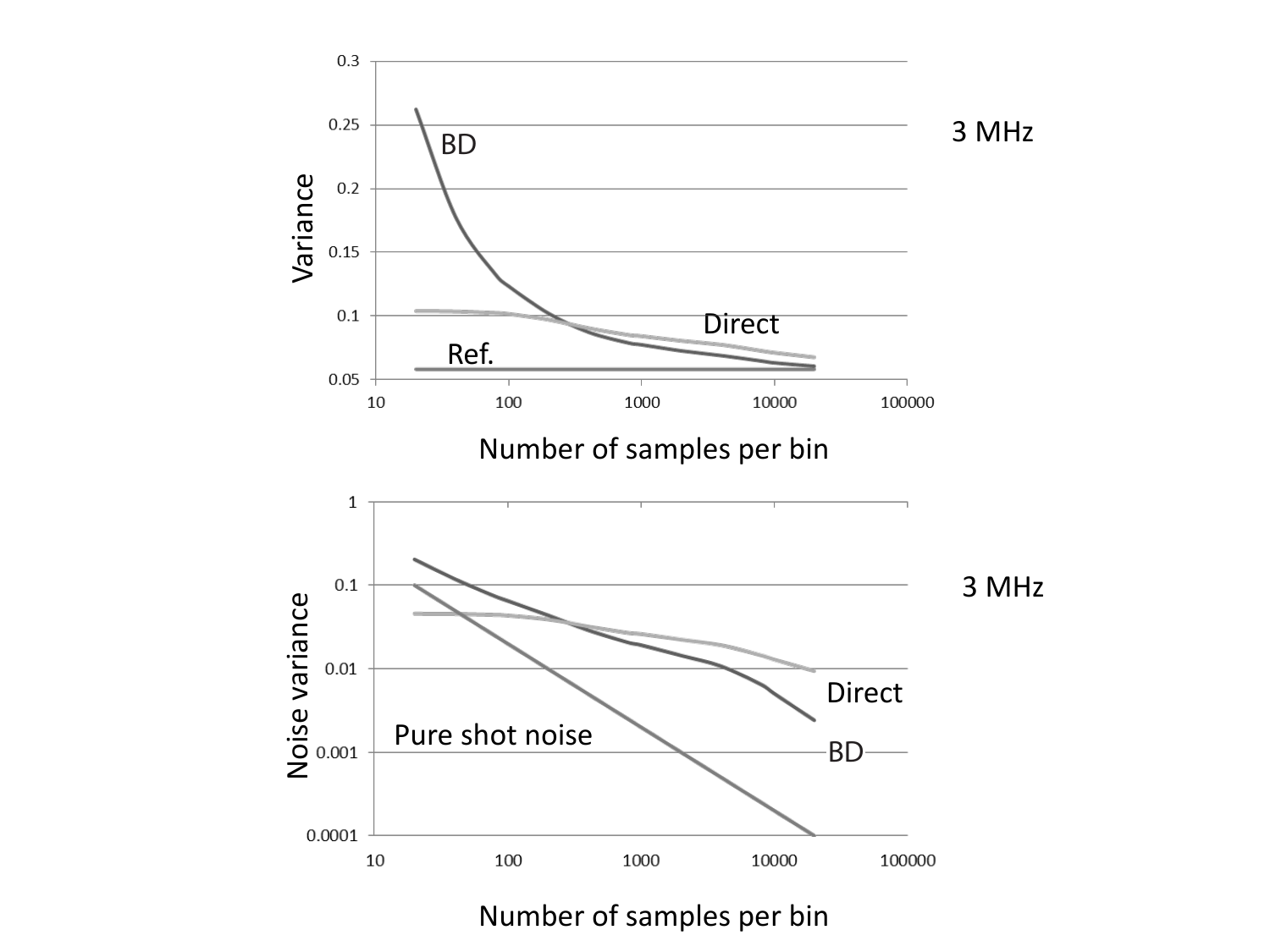}
 \caption{As in Figure \ref{var20MHz1}, except for an independent data set of
 960,000 samples acquired at the sampling interval of 0.167 $\mu$s (maximum
Fourier frequency of 3 MHz). Note how the reduction in the BD variance w.r.t.
the direct for large bin sizes is not as marked as in the 20 MHz sampling case
of Figures \ref{var20MHz1} and \ref{var20MHz2}.  Thus there seems to be an
advantage in sampling as much below the coherence time $\ta$ of the bunching
noise as possible.}
    \label{var3MHz}
\end{figure*}

\begin{figure*}[!h]
  \centering
    \includegraphics[angle=0,width=7in]{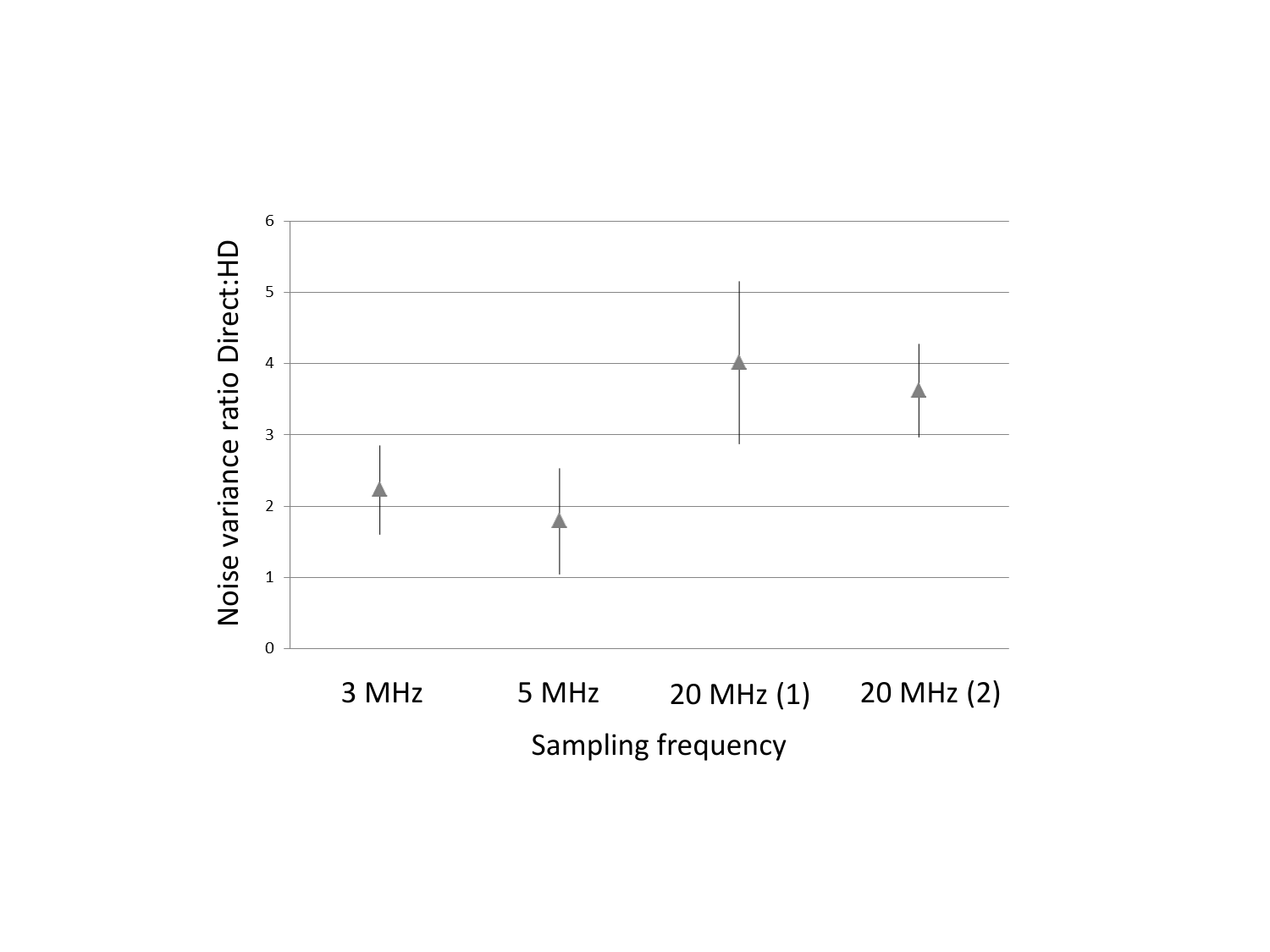}
 \caption{The ratio $r$ of the bunching noise variance, direct to BD,
 in the case when the time series is bin-averaged every 800 (\ie~$N=800$), as a
function of sampling frequency.
The error in the ratio is obtained by computing the variance in this ratio from
the data.
Quantum field theory \cite{zmu14,lie15,nai15} predicts $r=1$ for
large $N$ (the values of $N$ used in this graph satisfy the large $N$
condition), \ie~there should be no difference in sensitivity between the two
methods.  The graph is consistent with the partial reduction of the bunching
noise by the BD process first noted in Figure \ref{PedestalACFs}, but the
effect
is quite marginal (due to the large errors in $r$), especially towards lower
sampling frequencies.}
    \label{improve}
\end{figure*}

Next, we evaluate the covariance function of the time series, which in the case
of direct detection is closely related to the bunching noise autocorrelation
function often referred to as $g^{(2)}$.  At each of the four sampling
frequencies, the time series of the laser reference, direct, and BD
measurements
(the first two are shown in Figure \ref{pedestal} for the 3 MHz case) are
simultaneously compared in this test.

According to the theoretical prediction, the covariance to mean-squared ratios
of the direct and BD signals are, for stationary light, given by (\ref{covJ})
and (\ref{covD}).  And since $\nu^2 \ll 1$ by (\ref{nu}), this ratio for the
square BD time series (\ie~the decontaminated measurement of $D_j^2$) should,
by
(\ref{covD}), have (a) a sharp central spike of magnitude $\ap 2$, dropping
rapidly to assume (b) the Gaussian function of the classical bunching noise in
the direct current autocorrelation. We emphasize that we actually never
measured
a clean Gaussian in either the covariance function of the direct or squared BD
time series, but a central broad peak and a long tail because of the existence
of long range correlations in the bunching noise of the partially incoherent
light we used (thus $\ta$ is really the minimum coherence scale among multiple
ones).  In fact, such correlations prevent the noise-to-signal ratio from
assuming the simple $\propto \ta/(NT)$ form of the radiometer equation in
(\ref{errordirect}) for large exposure times $NT \gg \ta$, because the presence
of bunching noise components with timescales $> \ta$ modify the asymptotic form
of radiation.

Next, all bunching noise components are expected to affect the BD time series
in the same way as the direct, and so the first term on the right side of
(\ref{varDD1}) should still be modified in the same way as the last term of
(\ref{errordirect}). Any deviation of the BD noise-to-signal ratio from the
direct ratio would imply a possible enhancement or suppression of bunching
noise
in the BD process, since the $2(1+\nu^2)/N$ term in (\ref{varDD1}) is
negligible
in the limit of large $N$.

Lastly, although the light signal we used is explicitly time varying, {\it
viz.}
it is periodic (having the pedestal time series of Figure \ref{pedestal} due to
varying the laser output power), this non-stationarity occurs only on very long
timescales and does not distort the small lag features of the covariance
function on short timescales; rather, it only further extends the (already
long)
tail of the covariance function beyond that of the classical bunching noise of
stationary light.

Turning to the graphs of Figure \ref{PedestalACFs}, it should be stated upfront
that the sharp central spike referred to as (a) in the previous paragraph was
{\it always} detected with the correct height in the squared BD time series
({\it viz.} the filtered measurement of $D_j^2$), and because this shot noise
effect is insignificant when many samples are averaged, we will not show the
spike.
Instead, by avoiding the central two bins of the covariance data, we
could reveal much more clearly what really matters, which is the effect listed
as item (b) in the previous paragraph, {\it viz.} the shorter but much broader
peak of the classical bunching noise.  In this respect we shall find (below)
that apart from the ignored small lags, the covariance to mean-squared ratio of
the BD signal lies beneath the direct, but the difference becomes marginalized
as one moves to lower sampling frequencies. Higher sampling frequency data seem
to exhibit a greater difference between the direct and BD results (in terms of
the latter involving a smaller bunching noise amplitude).

In Figure \ref{PedestalACFs} we show the behavior in two such frequencies, 3
and
20 MHz, where the covariance to mean-squared ratio of the direct, BD squared,
and source (laser) reference fluxes are all seen.  Note the manner in which the
residual BD shot noise spill-over at small lags quickly gives way to the
bunching noise correlation (which is absent in the coherent laser signal, the
long covariance tail there is entirely due to the periodic variation of the
intrinsic laser power -- the pedestal of Figure \ref{pedestal} -- recalling
that
even 30,000 samples covers only 3/4 of a single laser power `step' within which
the laser intensity is held constant).

For the direct and BD detected incoherent light of the other two traces,
however, this correlation power is smaller in the case of the BD than the
direct.  In fact, the covariance tail of the BD merges with the intrinsic effect
of the reference signal's variation ahead of the direct.  These effects,
related to the reduced bunching noise in the BD time series mentioned in the
end
of the last section, become less marked as the sampling frequency increase to 5
MHz (which is still higher than the bunching noise frequency limit),
\ie~overall
we have good agreement between the two methods of detection, as expected by
theory.
Note also that the width of the bunching noise peak, {\it viz.} the tail after
the initial drop, is less for the 20 MHz experiment than the 5 MHz, indicating
that the bunching noise coherence length is very crudely a constant
\footnote{This is unlike the period of the reference signal variation, which is
only constant in terms of the number of samples, \ie~240,000 samples in six
steps, after which the laser power is recycled to its original value in the
next
experiment.} independent of the timing resolution of the data.

The performance of the direct and BD methods must further be compared in terms
of the variance for a given sample size $N$, \ie~(\ref{errordirect}) and
(\ref{varDD1}). This figure of merit is computed by bin-averaging the time
series data, using a fixed $N$ per bin $\cT$.
The results are shown in Figures \ref{var20MHz1} to \ref{var3MHz}, where we see
that the BD method appears to do better by this metric.  In section 3.2 we already discussed a possible reason for this, {\it viz.} the DD method suffers from nonlinearity problems.  Even though the magnitude seems too small to account for the anomaly, we do not have any direct data to exclude it.

\section{Discussion and conclusion}

We carried out a detailed experimental investigation to compare the performance
of direct versus BD measurement of the flux of incoherent light, with the
latter
inferring the flux from the shot noise variance of the light.  Despite the potential problems of detector non-linearity (section 3.2), this is to the best of the authors'
knowledge the first time that the bunching noise effect of high
occupation number chaotic light via the shot noise of the field has successfully been
measured, to the point of using it to infer the flux of the field. Furthermore, our
data indicate that apart from the additional bunching noise component, the
variance of the output of the BD deviates from the true (ensemble) flux only by
virtue of the shot noise inherent to the BD subtraction.

The BD method was qualitatively found to involve the same level and pattern of
bunching noise as the direct, in agreement with prior theoretical work
\cite{lie15,nai15,zmu14}. However,  some BD measurements (20 MHz data) seemed
to
indicate a smaller noise amplitude in terms of photon bunching, whilst
maintaining the noise pattern. This could be due to non-linearity in the DD process.  As explained in section 3.2, even
though simulations showed this should not be a problem, the conclusion was not quantitatively confirmed by measurements.


\appendix

\section{APPENDIX: Fluctuation of partially incoherent light}

\subsection{Directly detected signal}

Here we derive the key formulae of section 2B (for the ones in section 2A,
see\cite{lie15}). The partially incoherent light may be modeled as a mixture of
a coherent beam and chaotic beam.  So long as we are dealing with a narrow
bandwidth, it is more convenient to work with the Fourier transforms of the
annihilation and creation operators,
 \beq \ha(t) = \fr{1}{\sqrt{2\pi}} \int d\om\,\ha(\om)e^{-i\om t};
 \had(t) = \fr{1}{\sqrt{2\pi}} \int d\om\,\had(\om)e^{i\om t}. \eeq
They satisfy the commutation relations
 \beq [\ha(t), \ha(t')] = \de(t-t'). \label{comm} \eeq
The intensity is $\om_0 \had(t)\ha(t)$, but in the narrow-band case, it is
simpler to remove the factor of $\om_0$, and talk instead about
 \beq \hJ(t) = \had(t)\ha(t), \label{Jaa} \eeq
which represents the number of photons arriving per unit time.
For the coherent state we assign the mean photon arrival rate (or expectation
value of the same) as
 \beq \<\hJ(t)\>_1=  (1-\nu) \fr{n_0}{\ta}, \label{alpha} \eeq
and for the chaotic state
 \beq \<\hJ(t)\>_2 = = \nu \fr{n_0}{\ta}. \label{beta} \eeq
The mean rate of the of the full beam,  \beq \<\hJ(t)\> =  \<\hJ(t)\>_1 +
\<\hJ(t)\>_2 = \fr{n_0}{\ta},  \label{SUM} \eeq
since intensity is additive.

Next, we evaluate the covariance of the intensity time series
 \beq \cov(J(t), J(t')) = \<\hJ(t) \hJ(t')\> - \<\hJ\>^2, \label{covS} \eeq
where
\begin{eqnarray}
\<\hJ(t) \hJ(t')\> & = & \<\had(t)\ha(t)\had(t')\ha(t')\> \nonumber \\
                   & = & \<\had(t)\ha(t')\> \delta (t-t') +
                      \<\had(t)\had(t')\ha(t)\ha(t')\> .  \label{twopt}
\end{eqnarray}
For the coherent state the expectation value of the two point function is
\beq \<\hJ(t) \hJ(t')\>_1 = (1-\nu) \fr{n_0}{\ta}\delta (t-t') +
(1-\nu^2)\fr{n_0}{\ta}. \label{twoptC} \eeq
Thus the covariance is
\beq \cov(J(t), J(t'))_1 = (1-\nu) \fr{n_0}{\ta}\delta (t-t') \label{covC} \eeq
which clearly shows the coherent state has only shot noise and no bunching
noise.
Repeating the exercise to the chaotic state, assuming it is Gaussian
thermal light the last term in (\ref{twopt}) is expressible as
\begin{eqnarray}
\<\had(t)\had(t')\ha(t)\ha(t')\> & = & \<\had(t)\ha(t)\>\<\had(t')\had(t')\>
\nonumber \\
 & &+ \<\had(t)\ha(t')\>\<\had(t')\had(t)\>.
\label{GTlight}
\end{eqnarray}
As a result
\beq \cov(J(t), J(t'))_2 = \nu \fr{n_0}{\ta}\delta (t-t') + \nu^2
\fr{n_0^2}{\ta^2} e^{-(t-t')^2/\ta^2}. \label{covI} \eeq
Since the two constituent beams $1$ and $2$ are uncorrelated, the covariance of
the full beam is the just the sum of the two covariances, {\it viz.}
\beq \cov(J(t), J(t')) = \fr{n_0}{\ta}\delta (t-t') + \nu^2 \fr{n_0^2}{\ta^2}
e^{-(t-t')^2/\ta^2}. \label{covF} \eeq
This is the derivation of the complete version of (\ref{covJ}) in section 2B
which includes the shot noise contribution, {\it viz.} the first term on the
right side of (\ref{covF}), as well as the bunching noise.  By combining
(\ref{covF}) with (\ref{covS}) and (\ref{twopt}), we arrive at
\beq \<\had(t)\had(t')\ha(t)\ha(t')\> = \fr{n_0^2}{\ta^2} (1+\nu^2
e^{-(t-t')^2/\ta^2}), \label{fourpt} \eeq
an equation we shall find useful in due course.

If we define the average total flux over some time interval $T$ as
 \beq \hJ_T(t) = \fr{1}{T} \int_{t-T}^t dt'\,\hJ(t'), \label{JT} \eeq
then we find
 \beq \var(J_T(t)) =  \fr{1}{\ta T}\left[\nu^2 n_0^2 F\left(\fr{T}{\ta}\right)
 + n_0\right], \eeq
where
\beq F\left(\fr{T}{\ta}\right) = \fr{\ta}{T} \int_{-T}^T dt\,(T-|t|)|f(t)|^2. \label{FT} \eeq
Note that if $T\ll \ta$, the function $f(t)$ in the integrand will be reduced to $1/\ta^2$, so $F(T/\ta)\approx T/\ta$.
The relative uncertainty in the measurement of
$J_T$ is given by
 \beq \fr{\var(J_T(t))}{\<\hJ_T(t)\>^2}
 = \fr{\ta}{T} \left[\nu^2 F\left(\fr{T}{\ta}\right)
 + \fr{1}{n_0}\right], \eeq
or
 \beq \fr{\var(J_T(t))}{\<\hJ_T(t)\>^2} \approx \nu^2
 + \fr{\ta}{n_0 T},\ \ \textrm{for}\quad T\ll\ta
  \label{errorJ}\eeq
which agrees with (\ref{varS})
if we ignore the shot noise contribution of the last term.
On the other hand, if we measure for a much longer time $\cT=NT$,
we must replace $T$ in (\ref{JT}) by $\cT$ and use the limiting value of $F(x)$
for $x\gg 1$, {\it viz} $\sqrt{\pi}$.
Then we have, dropping the shot noise term,
 \beq \fr{\var(J_\cT(t))}{\<\hJ_\cT(t)\>^2}
 \approx \sqrt{\pi} \fr{\nu^2\ta}{\cT}=\sqrt{\pi} \fr{\nu^2\ta}{NT},
 \ \ \textrm{for}\quad \cT\gg\ta.
 \label{errorD}\eeq
and this is the derivation of (\ref{errordirect}).

\subsection{Difference signal for split beam}

In a 50:50 beam splitter, it is useful to consider a second input beam,
which is in fact in its vacuum state.  Let us represent the annihilation and
creation operators of that second input by $\hb(t), \hbd(t)$.  Then for the two
output beams we have annihilation operators
 \beq \hc = \fr{1}{\sqrt{2}}(\ha+i\hb),\qquad \hd = \fr{1}{\sqrt{2}}(\ha-i\hb).
 \label{split} \eeq
Note that $\hc$ and $\hd$ each satisfy the commutation relations (\ref{comm}),
and also (\ref{Jaa}) and (\ref{SUM}) with $n_0$ replaced by $n_0/2$ in the
latter.  Moreover, $\big[\hc,\hdd]=0$.

One might perhaps wonder if using $\hb(t)$ rather than $\hb(\om)$,
with the replacement of factors of $\om$ by $\om_0$, which is justified for the
narrow-bandwidth case, might be inadmissible for the vacuum contribution.
However, if one retains the factors of $\om$, they will be converted to time
derivatives that will ultimately act on other factors that are limited by
bandwidth, and the leading contributions will be given quite accurately by the
replacement of $\om$ by $\om_0$, so should not be a serious problem.

The quantity we are particularly interested in is the difference signal,
the difference between the numbers of photons arriving in the two output
channels.  This is given by
 \beq \hD(t) = \hcd(t)\hc(t) - \hdd(t)\hd(t). \eeq
Substituting from (\ref{split}) we see that this quantity may be written as
 \beq \hD(t) = i\had(t)\hb(t)-i\hbd(t)\ha(t). \label{Dab} \eeq
Obviously, its expectation value is zero:
 \beq \<\hD(t)\>=0. \eeq
The factorization between $\ha$ and $\hb$ operators makes this a very
convenient form to use.
For example, in computing the two-time function, we see that
 \beq \<\hD(t) \hD(t')\> = \<\had(t)\ha(t')\>\<\hb(t)\hbd(t')\>
+\<\ha(t)\had(t')\>\<\hbd(t)\hb(t')\>, \eeq
and because the $b$ input is in its vacuum state, the second term vanishes,
while in the first, $\<\hb(t)\hbd(t')\>=\de(t-t')$.  Thus we find
 \beq \cov(D(t),D(t')) = \<\hD(t) \hD(t')\> = \fr{n_0}{\ta}\de(t-t'). \eeq
So the measurement of the variance of $D$ provides a way of measuring $n_0$.

Of course, any measurement will take up a finite time interval.
We suppose as before that the total available time $\cT$ is divided up into $N$
small segments of duration $T$, and define the average flux in the $j$th
interval as
 \beq \hD_j = \fr{1}{T}\int_{(j-1)T}^{jT} dt\,\hD(t), \eeq
where we assume $T\ll \ta$, so that
 \beq \var(D_j) = \fr{n_0}{T\ta}, \qquad \cov(D_j,D_k)=0,\ (j\ne k). \eeq

Now to estimate the accuracy of the measurement we can make,
we need to compute the expectation value $\<\hD_j^2 \hD_k^2\>$.  However, for
use later we consider the more general case
\begin{eqnarray}
 \<\hD_j \hD_k \hD_l \hD_m\>
 & = & \fr{1}{T^4} \int_{(j-1)T}^{jT} dt_1
\int_{(k-1)T}^{kT} dt_2 \nonumber \\
& & \int_{(l-1)T}^{lT} dt_3 \int_{(m-1)T}^{mT} dt_4
 \<\hD(t_1)\hD(t_2)\hD(t_3)\hD(t_4)\>. \label{Dj4}
\end{eqnarray}
When we substitute from (\ref{Dab}),
each term in the expectation value can be written as a product of an
expectation
value of $\ha$ and $\had$ operators, and one of $\hb$ and $\hbd$ operators.
Moreover, the latter vanish if they have a $\hb$ on the right or a $\hbd$ on
the
left, and there must be equal numbers of each of the two terms in (\ref{Dab})
containing $\hb$ and $\hb^\dag$ operators.  So there are just two terms
remaining:
\begin{eqnarray}
 & & \<\hD(t_1)\hD(t_2)\hD(t_3) \hD (t_4)\> \nonumber \\
 & = & \<\had(t_1)\had(t_2)\ha(t_3)\ha(t_4)\>
 \times\<\hb(t_1)\hb(t_2)\hbd(t_3)\hbd(t_4)\> \nonumber \\
 & &
 + \<\had(t_1)\ha(t_2)\had(t_3)\ha(t_4)\>
 \times\<\hb(t_1)\hbd(t_2)\hb(t_3)\hbd(t_4)\>. \label{Dt4}
\end{eqnarray}
With the abbreviation $t_{jk}=t_j-t_k$, we see that
 \beq \<\hb(t_1)\hb(t_2)\hbd(t_3)\hbd(t_4)\>
 =\de(t_{13})\de(t_{24})+\de(t_{14})\de(t_{23}), \eeq
while
 \beq \<\hb(t_1)\hbd(t_2)\hb(t_3)\hbd(t_4)\> = \de(t_{12})\de(t_{34}), \eeq
so clearly the result will only be nonzero when the indices $(j,k,l,m)$
are equal in pairs.

Enlisting the last two equations and (\ref{fourpt}),
we see the first term on the right side of (\ref{Dt4}) is
\begin{eqnarray}
& &
\had(t_1)\had(t_2)\ha(t_3)\ha(t_4)\>\<\hb(t_1)\hb(t_2)\hbd(t_3)\hbd(t_4)\>
\nonumber \\
& = & \de (t_{13})\de (t_{24})\<\had(t_1)\had(t_2)\ha(t_1)\ha(t_2)\> +
\nonumber \\
& & ~~~\de (t_{14})\de (t_{23}) \<\had(t_1)\had(t_2)\ha(t_2)\ha(t_1)\>
\nonumber \\
& = & \fr{n_0^2}{\ta^2} (1+\nu^2 e^{-t_{12}^2/\ta^2})[\de (t_{13})\de
(t_{24})+ \de (t_{14})\de (t_{23})]; \label{first}
\end{eqnarray}
and the second term is
\begin{eqnarray}
& &
\<\had(t_1)\ha(t_2)\had(t_3)\ha(t_4)\>\<\hb(t_1)\hbd(t_2)\hb(t_3)\hbd(t_4)\>
\nonumber \\
& = & \<\had(t_1)\had(t_3)\ha(t_1)\ha(t_3)\> \de(t_{12})\de(t_{34}) +
\nonumber \\
& &
~~~\<\had(t_2)\ha(t_3)\>\de(t_{12})\de(t_{23})\de(t_{34})
\nonumber \\
& = & \fr{n_0^2}{\ta^2} (1+\nu^2 e^{-t_{13}^2/\ta^2})\de(t_{12})\de(t_{34})
+ \nonumber \\
& &
~~~\fr{n_0}{\ta}\de(t_{12})\de(t_{23})\de(t_{34}). \label{second}
\end{eqnarray}
Note the symmetry of this expression under permutations of $\{1,2,3,4\}$,
which results from the fact that the different $\hD_j$ operators commute with
each other.

Substituting (\ref{first}) and (\ref{second}) into (\ref{Dt4}),
and integrating over short time intervals, $T\ll\ta$ as in (\ref{Dj4}), we find
(bearing in mind the symmetry under the aforementioned permutations),
\begin{eqnarray}
 & & \<\hD_j \hD_k \hD_l \hD_m\> \nonumber \\
 & = &
 \de_{jk}\de_{lm}\fr{n_0^2}{T^2\ta^2}(1+\nu^2 e^{-t^2_{jl}/\ta^2})
 + \de_{jl}\de_{km}\fr{n_0^2}{T^2\ta^2}
(1+\nu^2 e^{-t^2_{jk}/\ta^2})+ \nonumber \\
& & \ \de_{jm}\de_{kl}\fr{n_0^2}{T^2\ta^2}
(1+\nu^2 e^{-t^2_{jk}/\ta^2})+
 \de_{jk}\de_{kl}\de_{lm} \fr{n_0}{T^3\ta}, \label{Djklm}
\end{eqnarray}
where $t_{jk}=(j-k)T$.

To obtain the covariance of $D_j^2$ and $D_l^2$, we set $k=j$ and $m=l$,
and remove the first of the seven terms in (\ref{Djklm}), since the term is
canceled by the product of expectation values.
This yields
 \bea \cov(D_j^2, D_l^2) = \fr{n_0^2}{T^2\ta^2}\nu^2 e^{-t_{jl}^2/\ta^2}
 + \fr{n_0^2}{T^2\ta^2} 2(1+\nu^2)\de_{jl}
 + \de_{jl}\fr{n_0}{T^3\ta}. \label{Dj2Dl2}\eea
This is the derivation of the complete version of (\ref{covD})
that includes the shot noise in the original (unsplit) beam as the last term of
(\ref{Dj2Dl2}).
The first term alone gives the covariance when $j\ne l$.
For $j=l$ we find
 \beq \var(D_j^2) = (2+3\nu^2) \fr{n_0^2}{T^2\ta^2} + \fr{n_0}{T^3\ta}. \eeq
Thus the fractional error is given by
 \beq \fr{\var(D_j^2)}{\<\hD_j^2\>^2}= (2+3\nu^2) + \fr{\ta}{n_0 T}. \eeq
This is the derivation of the full version of (\ref{varD})
that includes the shot noise variance $\ta/(n_0 T)$ in the original (unsplit)
beam as well.

Of course, as before we can do better by observing
for a longer time and computing the sample mean
 \beq \overline{D^2} = \fr{1}{N}\sum_{j=1}^N D_j^2. \eeq
Clearly,
 \beq \var(\overline{D^2}) = \fr{1}{N^2}\sum_{j,l=1}^{N} \cov(D_j^2,D_l^2). \eeq
When we substitute from (\ref{Dj2Dl2}), in the first term,
we can convert the sum over $j-l$ to a Gaussian integral:
 \beq \sum_j e^{-j^2T^2/\ta^2}
 \approx \fr{1}{T}\int dt\,e^{-t^2/\ta^2} = \fr{\sqrt{\pi}\ta}{T}. \eeq
Thus we obtain
 \beq \fr{\var(\overline{D^2})}{\<\hD_j^2\>^2} =
 \fr{1}{N}\left[\fr{\nu^2\sqrt{\pi}\ta}{T}
 +2(1+\nu^2) +\fr{\ta}{n_0 T}\right]. \eeq
This is the derivation of the full version of (\ref{varDD1})
that includes the shot noise variance in the original (unsplit) beam as the
last term.

\end{document}